# Silver plasmonic density tuned polarity switching and anomalous behaviour of high performance self-powered *β*-gallium oxide solar-blind photodetector

Kanika Arora, Vishal Kumar, Mukesh Kumar*

Functional and Renewable Energy Materials Laboratory, Indian Institute of Technology Ropar, Punjab, 140001, India

## ABSTRACT

Deep understanding of plasmonic nanoparticles (PNPs)-light interaction over semiconductors surface shows great promises in enhancing their optoelectronic devices efficiency beyond the conventional limit. However, PNP-light interaction critically decided by the distribution density of PNPs over the semiconductor surface which is not entirely understood. Here, a systematic study depicting how the interparticle gap between Silver (Ag) NPs influences the performance of the β-$Ga_2O_3$ based solar-blind photodetector. Interestingly, a remarkable transition is observed, where the varied interparticle gap not only changes the polarity but also reverses the traditional photodetector behaviour. The positive transient response of bare β-$Ga_2O_3$ photodetector with feeble DUV light switches its behaviour remarkably to 20 times enhance negative-photoresponse when decorated by sparsely-spaced Ag-PNPs with ultra-high responsivity of 107.47 A/W at moderate power and an incredible report-highest responsivity of 4.29 mA/W on single semiconducting β-$Ga_2O_3$ layer at self-powered mode. Moreover, as the density of the Ag-PNPs was further increased, the photocurrent decreases with illumination which dynamically reverses the traditional photodetector to unnatural anomalous effect. In particular, our study represents the first demonstration of plasmonic tuning effect to two active dynamic switching modes; i.e. reverse switchable and anomalous behaviour, the fundamentals of which have not studied experimentally yet. Finally, we propose a unified well-explained model to rationalize all observed experimental trends while set-up fundamental basis for establishing potential applications.

# INTRODUCTION

Solar-blind deep ultraviolet (DUV) photodetectors which response only in < 290 nm of incident light are attracting considerable research interests owing to their usage in numerous applications which includes ozone layer monitoring, missile launching and tracking detection, space and astronomical research, short range communications, security and many more [1]. These photodetectors endow high yield to noise ratio even operating under sun's intense illumination [2]. Recently, to meet the developing trend of "internet of things'' with the aim to make energy efficient, compact, multi-tasking, self-driven and response to feeble DUV signal has become main concern. Thus far, solar-blind DUV photodetectors have been fabricated using various tuned and wide bandgap materials like AlGaN, MgZnO, diamond etc. In consequence of the fact that these materials require alloying and have various associated defects, they offer high leakage current [1]. $β$-$Ga_2O_3$ is a promising material with a direct bandgap of ~4.9 eV and have magnificent intrinsic solar-blind nature that eradicate alloying and cost issue like problems. Moreover, it has excellent chemical, mechanical and thermal stability, making it an attractive substitute for solar-blind photodetectors applications[3]. Despite of the recent progress on β-$Ga_2O_3$ based photodetectors, which are mainly attempting to reduce native defects and to enhance performance and sensitivity, little progress has been achieved to improve their fundamental properties beyond the conventional limits. Nevertheless, the sensitivity, responsivity and other critical parameters of any photodetector are mainly affected by the absorption coefficient which is scarcely affected by solely increasing the fundamental properties of the absorbing material. The advent of plasomic science that involves dispersal of metal NPs into dielectric has brought several significant benefits to the photodetector[4]. Recently, PNPs have been utilized to improve the performance of the photodetectors due to their ability to increase absorption coefficient beyond the diffraction limit of semiconductors via strong electric field through local surface plasmon resonance (LSPR)[4,5]. Typically, the SPs can be realized by placing metal NPs on to the surface of the film, where the near field resonance of the plasmon by the incident photons makes collective oscillation of free excited electrons in metal nanoparticles and the scattering by the LSPR on semiconductor make more photons to reach the substrate while enhancing the overall device characteristics[6].

Although the PNPs promises high absorption and sensitivity but the free electrons motion near the surface of metals also generates heat by means of ohmic losses[7]. On the other hand, the extreme losses in the form of heat is been utilized by some applications such as steam generation [8], however it offers disadvantage for photodetector devices[7]. As a consequence, ambiguities between various structure parameters have caused conflicts regarding the usefulness of plasmonics in the literature that have inhibited the progressing of an integrated theoretical perception for their practical applications[7,9,10]. One highly effective approach towards achieving efficient photodetector is by modulating the density of plasmons distribution over semiconductor surface[11]. Till now the active control study of PNPs over semiconductor surface for optimum light harvesting capacity remains chiefly elusive and is crucial for many future practical applications[4].

Herein, a systematic study has been conducted investigating the dependence of optoelectronic properties of solar-blind β-$Ga_2O_3$ photodetector with varying density distribution of Ag PNPs over its surface. Interestingly, a remarkable transitions are found, where the varied distribution density of Ag PNPs changes the polarity and even reverses the traditional photodetector behaviour. According to the transient response of bare β-$Ga_2O_3$ photodetector, device exhibit positive photoresponse which switches its behaviour drastically 20 times enhanced negative photoresponse when decorated by sparsely placed 20 sec sputtered Ag PNPs (β-$Ga_2O_3$@Ag20s). Besides, β-$Ga_2O_3$@Ag20s device showed an ultra-high responsivity and an incredible report-highest responsivity of 4.29 mA/W on single semiconducting β-$Ga_2O_3$-layer at self-powered mode. In a follow-on study, by re-sputtering for another 20 s (β-$Ga_2O_3$@Ag40s) the bridging of Ag interparticle gap in aggregated form results in anomalous behaviour. The photocurrent reduced significantly upon light illumination, through plasmon light-to-heat conversion via photon-phonon coupling showing anomalous behaviour. Based on these facts, we construe that the charge carrier dominance process critically affected by the PNP distribution *via* a distinct mechanism in contrast by a conventional semiconductor-based photodetector. This opens new door for different set of applications for their admired ability to automatically switch off the system thus acting as a safe guard and other on-chip sensor[12-14].

# RESULTS AND DISCUSSION

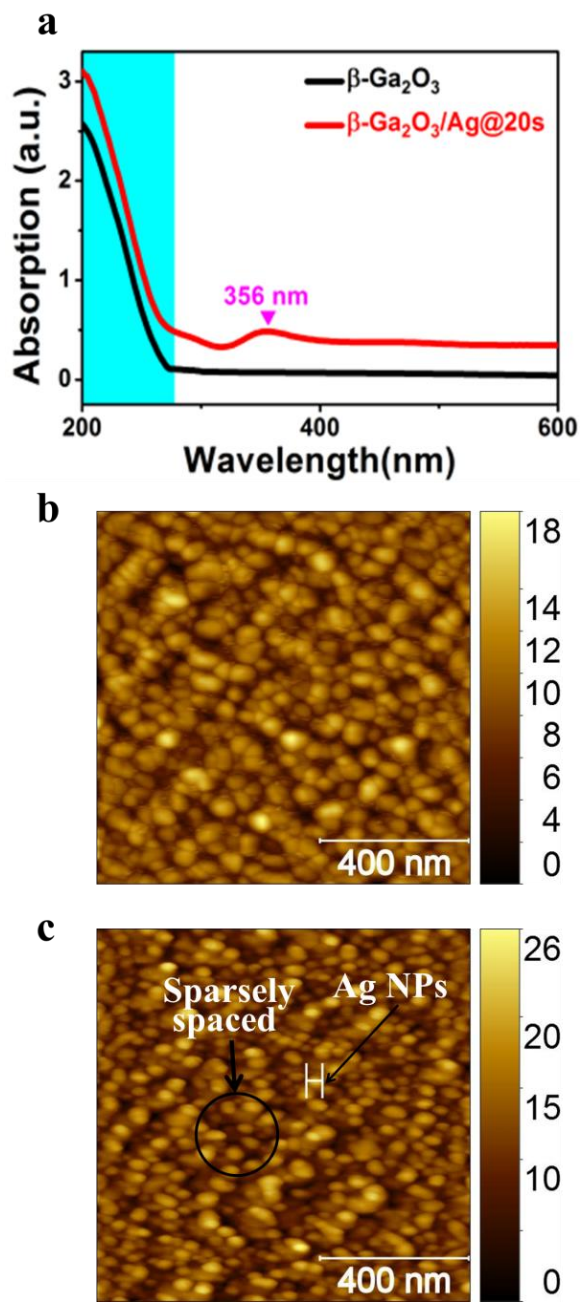

**Fig. 1. Optical absorption and AFM imaging of as-deposited and Ag decorated gallium oxide thin film. a** Absorption vs wavelength comparison of bare β-$Ga_2O_3$ and Ag NPs decorated β-$Ga_2O_3$/Ag@20 s. thin films **b** AFM measured surface morphology of as-deposited β-$Ga_2O_3$ thin film and **c** 20 sec Ag NPs decorated on β-$Ga_2O_3$ thin film. The encircle shows that the distribution of Ag NPs are sparsely spaced.

Fig. 1 illustrates the optical properties of as-deposited β-$Ga_2O_3$ and 20 sec sputtered Ag decorated β-$Ga_2O_3$, β-$Ga_2O_3$/Ag@20s, thin films. Fig. 1a clearly shows that optical absorption increases with introduction of Ag NPs over β-$Ga_2O_3$ thin film. β-$Ga_2O_3$ absorbs in deep solar-blind region as can be seen in absorption spectra. While, Ag NPs absorb radiation via LSPR and presence of an additional peak at 356 nm in absorption spectra confirms it. Normally, Ag NPs absorbs above 400 nm, the shift in absorption peak is owing to the change of dielectric medium around Ag nanoparticle [15]. We further analyze the topography of our samples, Fig. 1 (b and c) shows the AFM of deposited thin films. From Fig. 1c it can be clearly seen that Ag NPs are sparsely distributed all over the β-$Ga_2O_3$ surface. Furthermore, Ag NP size is also computed and it comes out to be approximately 37±3 nm. The root-mean-square roughness ($R_m$) for bare β-$Ga_2O_3$ thin film and β-$Ga_2O_3$/Ag@20s on a 1 $\mu m^2$ scan area is found to be 3.5±0.51 nm and 3.7±0.23 nm, respectively. The results reveals that with sputtering of Ag NPs, the roughness of the film increases. On the basis of the optical properties, we can say that the Ag NPs plays a significant role to increase in the absorption of a photon in gallium oxide thin film. For comparison, Fig. S1a, b (Supplementary Information) shows the optical absorption spectra and its zoomed version for bare β-$Ga_2O_3$, β-$Ga_2O_3$/Ag@20s and β-$Ga_2O_3$/Ag@40s samples. It can be clearly inferred that increase in deposition time of Ag NPs on gallium oxide surface leads to enhancement in optical absorption.

Fig. 2a shows the semi-log scale I-V spectra of Ag NPs (20s) decorated β-$Ga_2O_3$ thin film photodetector. This clearly shows that device have asymmetric rectifying characteristic behavior as intensity of negative photocurrent is very high as compared to intensity of positive photocurrent generated in the photodetector. Inset of Fig. 2a shows the semi-log scale I-V spectra of pristine β-$Ga_2O_3$ thin film based photodetector which, also shows the asymmetric rectifying characteristic. In contrast, for pristine β-$Ga_2O_3$ thin film based photodetector the intensity of positive photocurrent is higher than the negative photocurrent. The higher photocurrent at the negative side of the β-$Ga_2O_3$/Ag@20s device reflect that Ag NPs sparsely placed plays a major role in the conduction of hot electrons for the enhanced photocurrent in the negative bias side.

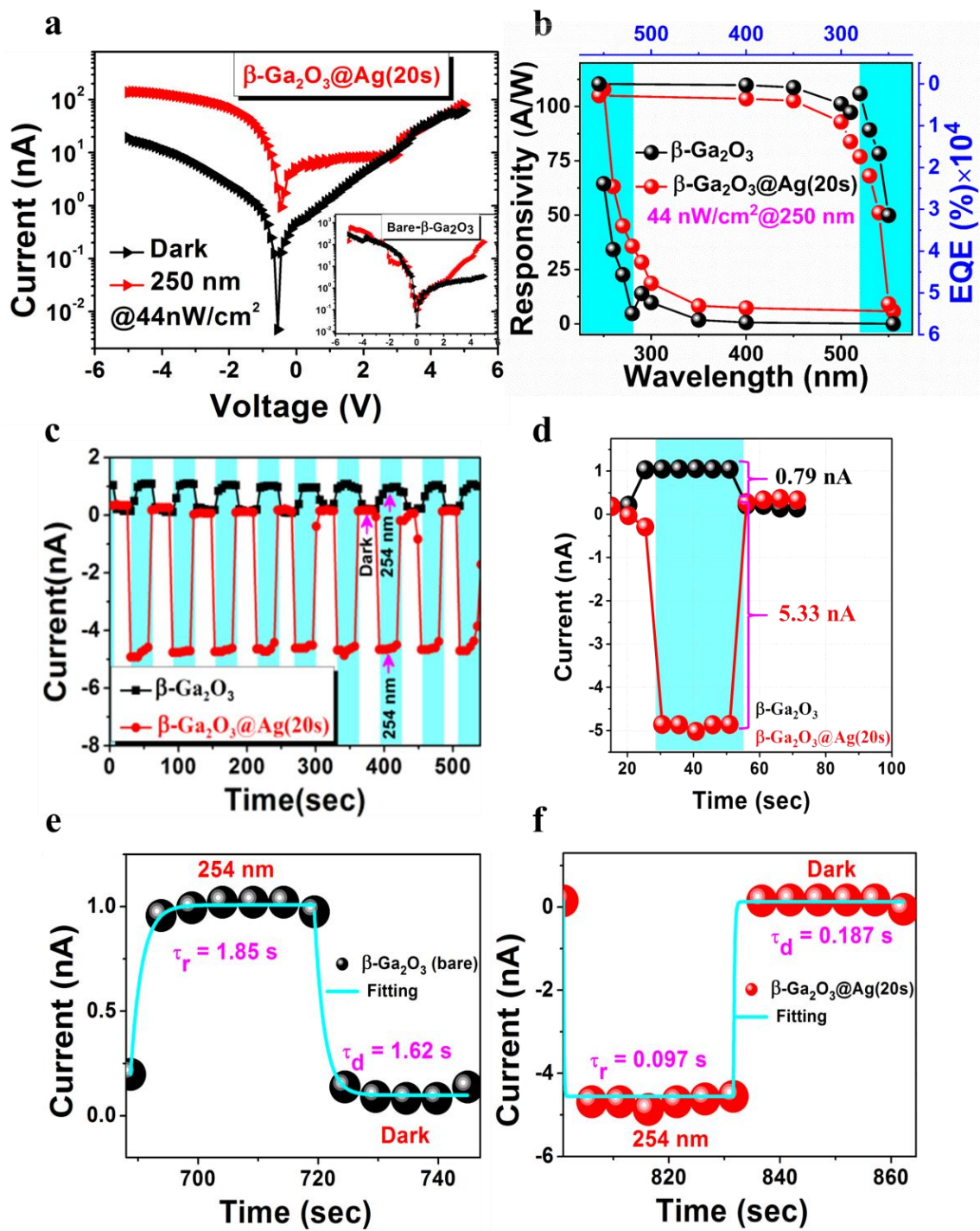

**Fig. 2. Optoelectronic performance of as-deposited and Ag decorated gallium oxide photodetectors. a** Semi-log scale I-V spectra of the photodetector under 250 nm illumination for β-Ga$_2$O$_3$/Ag@20s photodetector. Inset shows the I-V response of bare β-Ga$_2$O$_3$ photodetector and **b** Responsivity, External quantum efficiency (EQE), vs wavelength of bare β-Ga$_2$O$_3$ and β-Ga$_2$O$_3$/Ag@20s photodetector at feeble irradiation intensity of 44nW/cm$^2$. The Responsivity and EQE are enhanced for Ga$_2$O$_3$/Ag@20s. **c** Transient photo-response reproducibility at 0 bias under exposure to 254 nm illumination of bare β-Ga$_2$O$_3$ and β-Ga$_2$O$_3$/Ag@20s, **d** Single cycle of transient response comparison of bare β-Ga$_2$O$_3$ and β-Ga$_2$O$_3$/Ag@20s, **e, f** Fitting of the response time for bare β-Ga$_2$O$_3$ and β-Ga$_2$O$_3$/Ag@20s photodetector, respectively.

Fig. 2b displays the variation of computed responsivity of bare β-Ga$_2$O$_3$ sample and Ag NPs decorated β-Ga$_2$O$_3$/Ag@20s photodetector with respect to wavelength. It is clear that the introduction of Ag NPs over β-Ga$_2$O$_3$ thin film surface leads to the significant increase in the responsivity of the fabricated device as responsivity for bare β-Ga$_2$O$_3$ sample is 64.47 A/W which is enhanced to 107.86 A/W with the introduction of Ag NPs under 250 nm illumination. It should be noted that, all of these measurements are taken under ultra-low illumination intensity of 44 nW/cm$^2$, which is appreciated for many applications as discussed where detecting extremely feeble signal is at premium. A similar trend was perceived for external quantum efficiency (EQE) where $3.35 \times 10^4$% EQE for bare β-Ga$_2$O$_3$ got boosted to $5.35 \times 10^4$% with Ag NPs over the β-Ga$_2$O$_3$ surface. Apparently, the higher EQE for Ag coated β-Ga$_2$O$_3$ photodetector is due to the advantage of effective charge transfer between Ag plasmons and β-Ga$_2$O$_3$ surface. The performance of the β-Ga$_2$O$_3$/Ag@20s photodetector presented here in term of EQE and responsivity even beats our previous work [16].

Reproducible on-off I-T switching operating under self-biased condition is shown in Fig. 2c with 254 nm illumination. The turning point of the photodetector in photocurrent when decorated with Ag NPs can be clearly seen the transition from positive response of bare β-Ga$_2$O$_3$ photodetector to enhanced negative response of β-Ga$_2$O$_3$@Ag20s photodetector. This type of polarity photoconductance with a switching effect by means of NPs is uniquely found and the fundamental mechanism prevailing this phenomenon is discussed later on in detail. The observed polarity distinctive transient response with NPs is first time observed and reported in this paper. Fig. 2d shows single transient response comparison of β-Ga$_2$O$_3$ sample and Ag NPs decorated β-Ga$_2$O$_3$ sample. Thus, it is clear that introduction of plasmons (Ag plasmon) over the surface of β-Ga$_2$O$_3$ not only enhances the photocurrent ~6.66 times, but also completely reverses the photocurrent polarity leading to ambipolar behavior of photodetector. The enhancement in photocurrent intensity suggests that introduction of plasmon NPs over β-Ga$_2$O$_3$ surface increases the concentration of charge carriers significantly in β-Ga$_2$O$_3$. Here, we have evaluated the responsivity under self-biased condition and is found to be 0.637 mA/W for bare β-Ga$_2$O$_3$ photodetector and exceptional 4.29 mA/W for Ag decorated β-Ga$_2$O$_3$@Ag20s photodetector. Remarkably, the value of responsivity for β-Ga$_2$O$_3$@Ag20s photodetector is the highest reported for single β-Ga$_2$O$_3$ semiconducting based device. Thus it reflects the strong effect of Ag NPs for the enhanced responsivity of photodetector. For any good photodetector, along with high responsivity it must have least possible response time which tells us how fast the device responds to incoming signal. Thus, it is one of the major parameter deciding the performance of the photodetector. Response time of the photodetector was computed using equation (1).

$$I = I_o + Ae^{\frac{-t}{\tau_r}} + Be^{\frac{-t}{\tau_d}} \tag{1}$$

where $I_0$ is the steady-state photo-current, t denotes the time, A and B are constant factors, and $\tau$ represents the relaxation-time constant, $\tau_r$ and $\tau_d$ symbolize the rise and decay time constants, respectively. Values of rise time ($\tau_r$) and decay time ($\tau_d$) obtained for bare β-Ga$_2$O$_3$ and Ag decorated β-Ga$_2$O$_3$@Ag20s photodetectors by fitting the single cycle, Fig. 2 (e and f). Both the rise time and decay time for β-Ga$_2$O$_3$@Ag20s ($\tau_R$ = 0.097 s and $\tau_D$ = 0.187s) are significantly smaller than that obtained for the bare β-Ga$_2$O$_3$ thin film photodetector ($\tau_r$ = 1.85s and $\tau_d$ = 1.62s), indicating approximately 20 times faster response of UVC light signal for β-Ga$_2$O$_3$@Ag20s photodetector. These data facts clearly show that the introduction of Ag plasmon over β-Ga$_2$O$_3$ sample speed-up the carrier response in photodetector.

Comparison of our photodetector performance with gallium oxide heterostructure and non-gallium oxide based photodetector reported in literature with exceptional features is being specified in Table 1. Herein, we have compared essentially most important optoelectronic parameters like responsivity and response time operating in self-biased conditions that determine the performance of the device. Therefore, it will be of great importance if some procedure can be taken to enhance these parameters. Although at self-biased responsivity value of 0.09 mA W$^{-1}$ [17] and 0.01 mA W$^{-1}$ [2] have been reported for epitaxial β-Ga$_2$O$_3$ vertical Schottky detectors and β-Ga$_2$O$_3$-nanowire, respectively. However, the responsivity of our photodetector is greatly improved to the value of 4.29 mA/W with Ag NPs. This is the highest value ever reported for a single semiconductor layer β-Ga$_2$O$_3$ based photodetector. Besides, we not only get the upshot of the enhanced performance but also the symmetric negative switching effect by NP in our device. Adding on we introduced another salient feature of first ever reported polarity distinctive transient response with NPs.

The result shows that the performance of the β-Ga$_2$O$_3$/Ag@20s photodetector cross-over the highest reported values in literature for gallium oxide and even go across our recent work despite of using lower work function Ag electrodes in the present work[16]. The performance of β-Ga$_2$O$_3$/Ag@20s shows high performance in self-powered as-well as in moderate-powered modes in comparison to the reported literature.

**TABLE 1. Comparison of fabricated Ag decorated β-Ga₂O₃ photodetector with other self-powered photodetector**. The Ag decorated β-Ga₂O₃ photodetector shows high performance with exceptional feature of negative polarity-distinctive photo-response.

| Materials | Device type | $R_\lambda$ (A/W) | Response time | Comments | Ref. |
|---|---|---|---|---|---|
| ZnO nanowire array | Schottky | $1.82 \times 10^{-3}$ | 81/95 ms | Thermal tuning | 18 |
| ZnO nanorod | MIS | $1.78 \times 10^{-6}$ | 0.1/0.1 s | Piezotronic enhancement | 19 |
| Polyaniline/ MgZnO | organic–inorganic hybrid | $160 \times 10^{-6}$ | 4.8/5.1 s | - | 20 |
| Ga₂O₃/Au nanowire array | Schottky | $1 \times 10^{-5}$ | 1/64 µs | Ultrafast responsivity | 2 |
| Ga₂O₃/Ga:ZnO | Heterostructure | $0.76 \times 10^{-3}$ | 0.179 s | - | 21 |
| ZnO/ Ga₂O₃ | Heterostructure | $9.7 \times 10^{-3}$ | 0.1/0.9 ms | - | 22 |
| Ga₂O₃ | Schottky | $1.4 \times 10^{-3}$ | 1.9/0.5 s | - | 23 |
| Ga₂O₃/NSTO | Heterojunction | $2.6 \times 10^{-3}$ | 1.08/0.65s | - | 24 |
| Ga₂O₃/Ag | Schottky | $4.29 \times 10^{-3}$ | 0.097/0.187 s | Negative polarity-distinctive photo-response | This work |

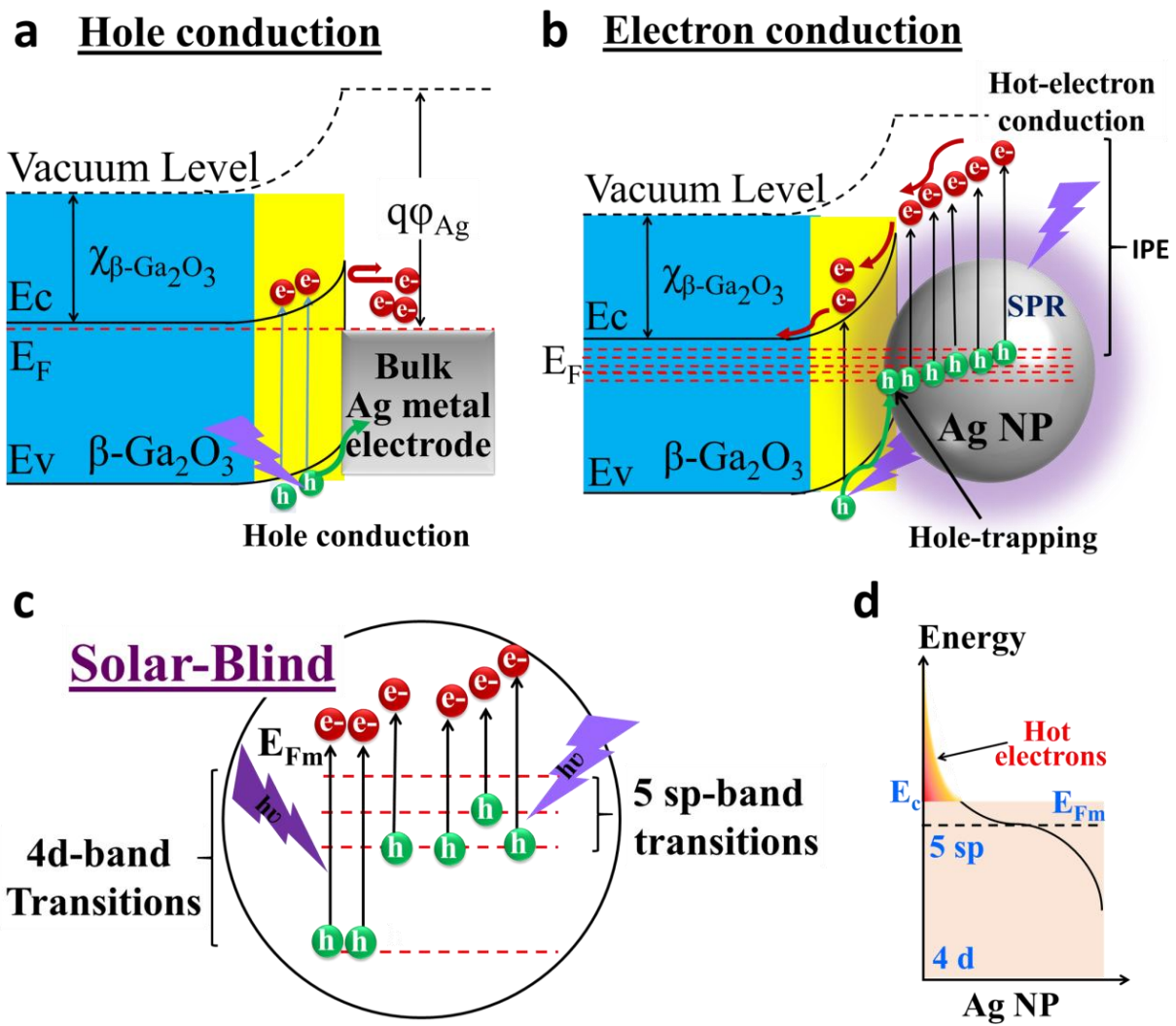

**Fig. 3**. **Mechanism study of pristine and Ag decorated gallium oxide photodetectors**. **a** Band alignment for bare β-$Ga_2O_3$ – bulk Ag metal electrode showing only hole conduction (positive conduction) for bare β-$Ga_2O_3$ photodetector **b** Band alignment of β-$Ga_2O_3$ with Ag nanoparticles in β-$Ga_2O_3$/Ag@20s showing hot electrons from Ag NPs surpassing the Schottky barrier into β-$Ga_2O_3$ and possible electron conduction (negative conduction) and **c** Schematic diagram of hot electron jumping from s and d orbitals of Ag NPs **d** Surface plasmon resonance excitation of Ag NP in the form of hot electrons, which momentarily occupy the conduction band empty states above the Fermi energy.

**Photocurrent in the photodetector without the Ag NPs** Fig. 3a describes the band alignment of the fabricated bare β-$Ga_2O_3$ photodetector device. When β-$Ga_2O_3$ is irradiated with 254 nm deep UV light, there is a generation of e/h pairs. Thus, electrons rise from its valence band to conduction band, leaving behind a hole. The electron affinity ($\chi$) of β-$Ga_2O_3$ is 4.0 eV, which is lower than the work function ($\varphi_M$) of our metal electrode Ag, 4.26 eV. So, Fermi level of semiconductor lies above the Fermi level of metal. Hence equilibrium will be reached by transportation of electrons from Fermi level of the semiconductor to that of metal, thus equalizing Fermi levels of both semiconductor and metal to the same level[25]. This implies metal side has an excess of electrons as compared to that of semiconductor which has an excess of holes. Thus, a net negative charge is developed at the metal side, while semiconductor develops a net positive charge, creating a Schottky barrier and it is clearly shown in Fig. 3a. Here we can see that there is a Schottky barrier formation of height $\varphi_B$ of 0.26 eV ($\varphi_B = \varphi_M - \chi$), thus inhibiting the motion of electrons from metal to semiconductor. But excess of holes remaining on semiconductor side are free to move, and they start mobilizing towards the metal side, thus generating a net positive photocurrent with UVC light illumination. Under these conditions if $p_1$ and $n_1$ are the total photo-generated holes and electrons when illumination is turned on and $p_2$ denotes the number of holes getting transported to metal side then net photocurrent generated can be expressed as:

$$\Delta I = I_p - I_d = A(\frac{-n_1}{n_1 + n} J_e + \frac{p_1 + p_2}{p_1 + p_2 + p} J_h) \qquad (2)$$

where n, p are original electron and hole carriers in β-Ga$_2$O$_3$ film, $J_e$ and $J_h$ are current densities of electrons and holes, $I_p$ and $I_d$ are currents under illumination and dark conditions and A is the electrically active area. Thus, above equation clearly suggests positive photocurrent generation under given constraints.

Furthermore, asymmetric behaviour of the formed photodetector device, as seen in Fig. 1a, can be explained by the dissimilarity in the source drain work functions that makes our device to work even in self-powered conditions. The observed dissimilarity may exist with the effect of difference in the point contacts present in the β-Ga$_2$O$_3$ thin film due to lattice ending periodicity contact and Ag metal contact at source and drain side. As indicated by our previous study, this can be attributed to the well-known fact of the difference in quasi-fermi pinning caused by the interfacial defects, adsorption on the surface and the resonant shallow level formation due to deep donor defects. We observed the asymmetric behaviour of I-V characteristics on both the photodetectors which clearly indicates dissimilar Schottky barrier height (SBH) at source and drain sides[16]. This facilitates the likelihood of realizing a good photo-conduction even at zero bias.

**Negative photocurrent in the photodetector with the Ag NPs** For β-Ga$_2$O$_3$/Ag NPs@20sec sample, the band alignment diagram is shown in Fig. 3b. It demonstrates a more comprehensive picture of the Ag NP-dependent negative photoresponse. Because the Ag NPs are sufficiently small and spherically shaped, therefore, under the illumination Rayleigh scattering bring into existence. This offers larger scattering angles and ensures momentum conservation at the β-Ga$_2$O$_3$/Ag NP interface to launch hot electrons via SPR decay. Before light illumination, the silver electron occupancy obeys the Fermi–Dirac distribution. However, when Ag NPs absorbs DUV irradiation, the electron Fermi-Dirac distribution function recognize both high energy hot electrons caused by the surface-assisted quantum transitions and Drude electrons, as shown in Fig. 3d. The energy of the DUV photons brings up the electrons above the Fermi level, giving rise to a population of energetic 'hot' electrons. This Ag NP surface-assisted transition of photons can be described in two ways: Radiative (by re-emission of photons) or non-radiative (by transference of energy in the form of hot electrons). Typically, non-radiative decay takes place either via intraband transitions (transitions between the partially filled 5sp bands to conduction band) or interband transitions (transitions between 4d to conduction band). In general the plasmonic-modulated/semiconductor systems allow for the transfer of only hot energetic electrons from the metal to the semiconductor by injecting above the Schottky barrier. Such energetic hot electrons can surmount the interface Schottky barrier if their energies are greater than the Schottky barrier height (SB), thus are able to migrate into the conduction band of the semiconductor. Soon after UVC absorption, the plasmon excitation decays energy of the hot electron conduction band which lie 4 eV above it, for Ag plasmon. Therefore, when irradiation of 254 nm (solar-blind) is incident, these hot electrons get injected from 4d and 5sp bands by crossing over the Schottky barrier height into the semiconductor thus giving rise to negative photocurrent. Here, owing to the small height of the Schottky barrier formed (0.26 eV), a large number of excited hot electrons cross over the Schottky barrier. Thus, the Ag NPs emits a huge amount of hot electrons that can be transferred to semiconductor side through internal photoemission (IPE) via non-radiative plasmon decay, thus increasing the contribution to the negative conductivity significantly. So, after the introduction of Ag plasmons over β-Ga$_2$O$_3$ if n$_2$ denotes the electron which gets transferred to semiconductor side, photocurrent generated can be given by:

$$\Delta I = A(\frac{-n_1 + n_2}{n_1 + n_2 + n}J_e + \frac{p_1}{p_1 + p}J_h) \qquad (3)$$

Equation 3 clearly shows negative photocurrent generation as electron carriers are greater in number as compared to hole carriers. Hence, the change in polarity of the generated photocurrent is thus explained as initially carriers (holes) were moving from semiconductor side to metal side however after the introduction of Ag NPs over β-Ga$_2$O$_3$ surface charge carriers (electrons) start moving from metal to semiconductor side. As, these Ag NP-semiconductor interfaces lie over the entire surface of the active region of our photodetector, it together leads to generate huge amount of negative photocurrent which has been observed in transient response of our photodetector.

**Interplay of the Ag NPs spacing effect in the photocurrent of the photodetector**

In follow up, to study how the spacing between plasmons changes the behavior of β-$Ga_2O_3$ based photodetector, Ag sputtering time was increased to additionally 20 secs. The typical topography was investigated by AFM and the results are shown in Fig. 4a. It can be clearly seen that Ag NPs are placed in tightly spaced agglomerated form overall on the surface of β-$Ga_2O_3$ thin films. To analyze the optoelectronic properties, current versus voltage of the β-$Ga_2O_3$/Ag@40s photodetector is taken and depicted in Fig. 4b. This figure shows that dark current has increased significantly with increasing Ag deposition time to 40 sec over β-$Ga_2O_3$ surface. Fig. 4c shows the reproducible I-T on-off switching for 40s Ag decorated β-$Ga_2O_3$ sample with 254 nm light source. Usually for light driven traditional photodetectors, photocurrent increases with shining light however, here a complete anomalous behavior is observed. Also, we can observe this peculiar behavior in highly repetitive manner where the photocurrent decreases with respect to the dark current with incident light on β-$Ga_2O_3$/Ag@40s photodetector. Concurrently, using the single cycle of transient response, response time for the photodetector is computed which can be seen in Fig. 4d. For this photodetector $\tau_r$ and $\tau_d$ computed out to be 0.34s and 1.84s respectively. This anomalous behavior of photodetector has been rarely reported and the reason behind this phenomenon is still under debate, which nevertheless have many potential applications. We have investigated the fundamental basis of this type of prevailing phenomena first time, the details discussed later.

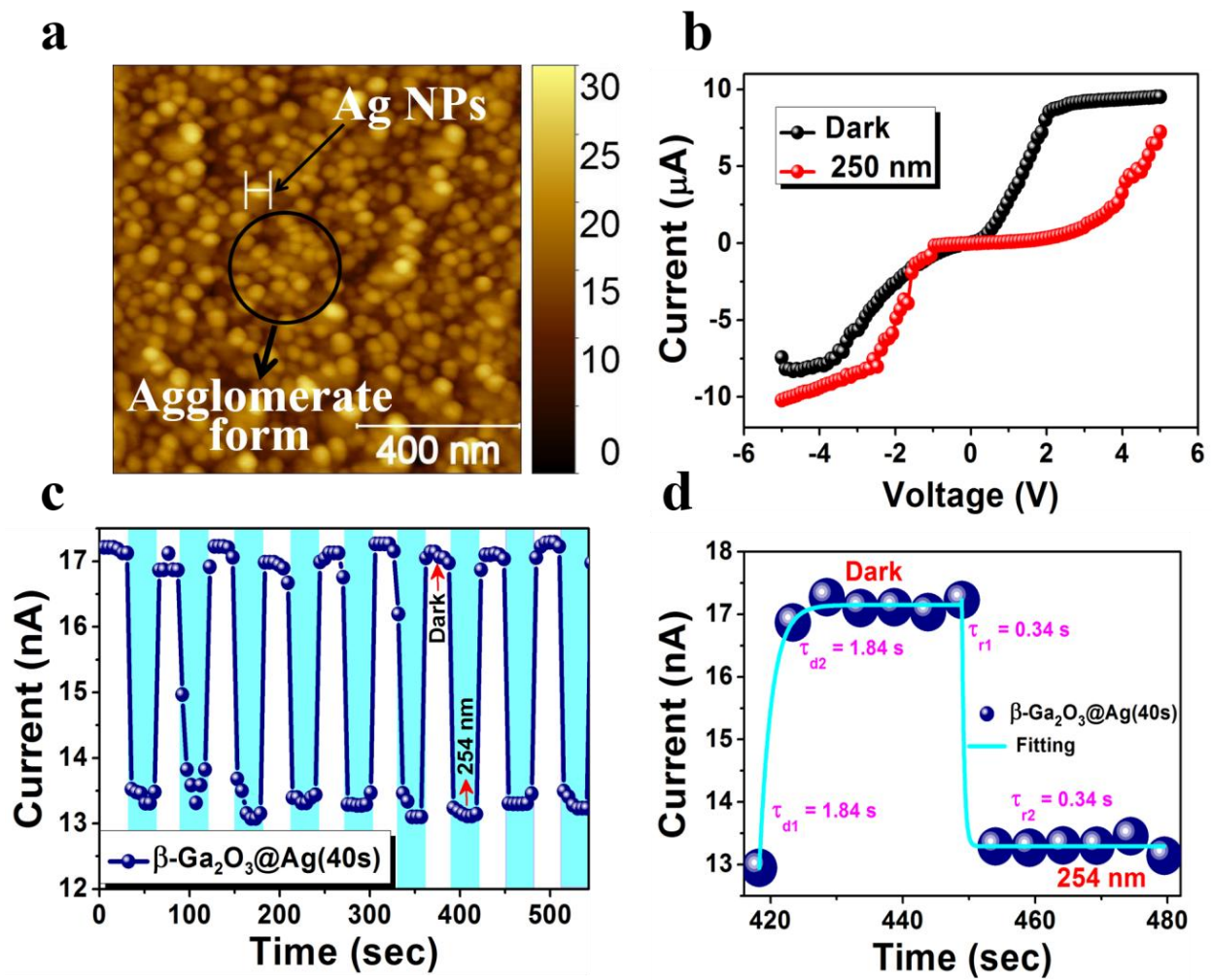

**Fig. 4. AFM imaging and optoelectronic performance of Ag decorated gallium oxide thin film. a** Surface topography of 40 sec sputtered Ag on β-$Ga_2O_3$ thin film. Encircled part shows the agglomerated form of Ag NPs distributed all over the β-$Ga_2O_3$ surface, **b** I-V of the fabricated device under 250 nm illumination for Ag NPs decorated sample, **c** Reproducible on-off I-T switching and **d** response time for β-$Ga_2O_3$/Ag@40s photodetector

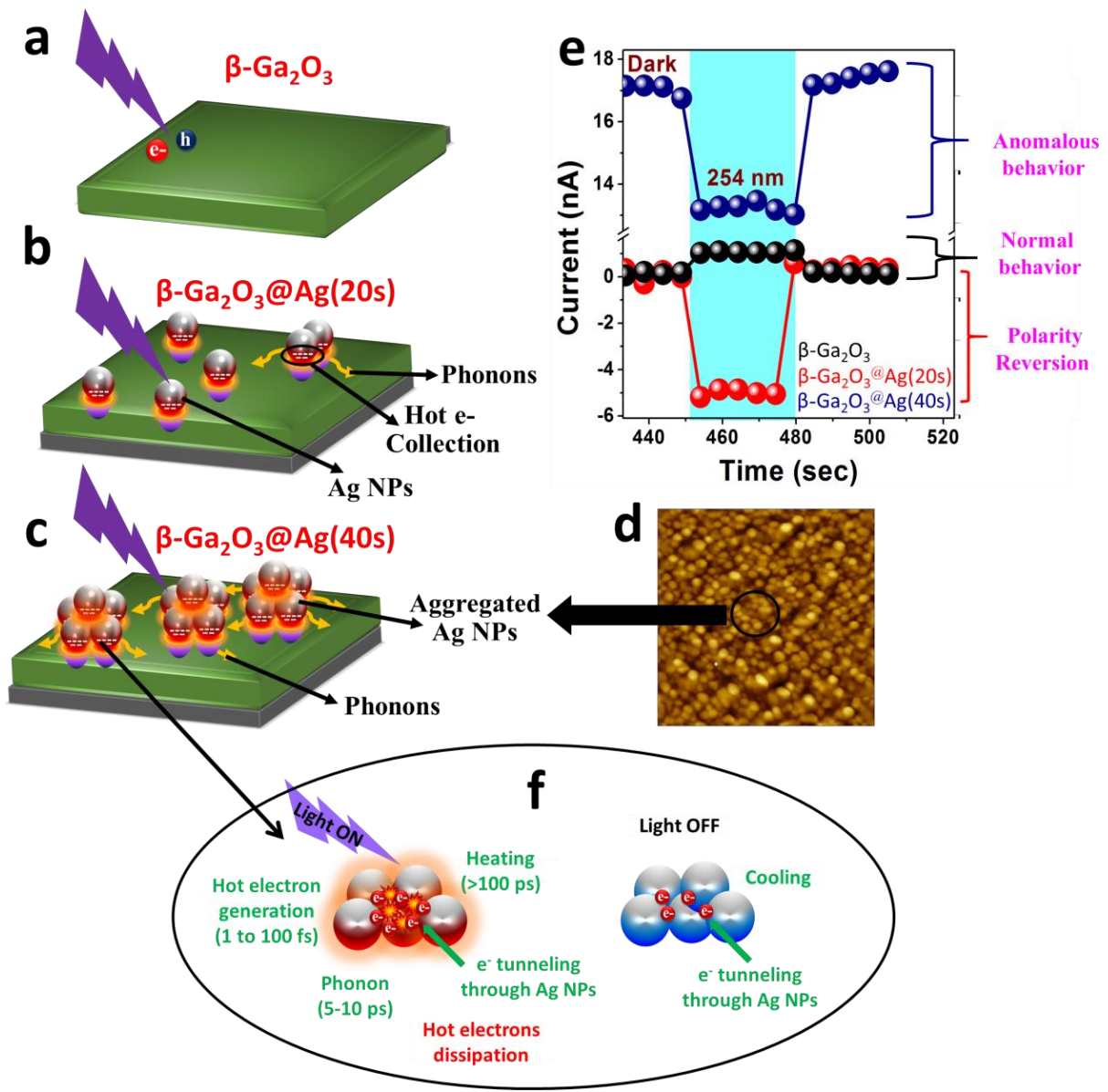

**Fig. 5. Unified schematic model for all observed experimental trends. a** to **c** The schematic shows the interaction of 254 nm light with bare β-$Ga_2O_3$, β-$Ga_2O_3$/Ag@20s and β-$Ga_2O_3$/Ag@40s photodetectors. The interaction of light with β-$Ga_2O_3$/Ag@20s and β-$Ga_2O_3$/Ag@40s leads to the generation of hot electrons followed by heat dissipation from metal nanoparticle to β-$Ga_2O_3$ via phonon. **d** Surface topography of β-$Ga_2O_3$/Ag@40s sample clearly shows that the Ag nanoparticle agglomerate to each other. **e** The comparison of photocurrent from bare β-$Ga_2O_3$, β-$Ga_2O_3$/Ag@20s and β-$Ga_2O_3$/Ag@40s photodetector. The results clearly reveals the effect of Ag nanoparticle on β-$Ga_2O_3$ to control the photoresponse. **f** Schematic of the Ag plasmonic light-to-heat photothermal conversion through photon–electron–phonon couplings. When the light is turned off, the cooling can be achieved by decaying the heat dissipation via equalizing lattice temperature.

Fig. 5 describes how the variation of Ag sputtering time leads to the variation in optoelectronic properties of fabricated photodetector. For bare β-$Ga_2O_3$ photodetector the incident DUV light results in generation of electron-hole pair in β-$Ga_2O_3$ semiconductor as shown in Fig. 5a.

However, deposition of the silver metal PNPs (Fig. 5b and 5c) results in strongly confined fields with hot electron generation in Ag NPs through absorption of DUV photon under the time scale of 1-100 femtoseconds. These generated hot electrons get transferred to β-Ga$_2$O$_3$ film by crossing over the Schottky barrier forming between the interface of Ag NPs and β-Ga$_2$O$_3$ semiconductor. This increases the optical absorption rate in β-Ga$_2$O$_3$ photodetector, Fig. 1a, and contribute to enhanced photocurrent, Fig. 2a. The experimental and theoretical model predicts; the lifetime of hot electron carriers exceeds the picoseconds timescale by transferring over the barrier height of metal/semiconductor interface.

The AFM of 40 s sputtered gallium oxide surface, Fig. 5d, clearly reveals that the Ag nanoparticle are in denser agglomerated form. In the Fig. 5e we have compared the single transient response, which shows how photo-response behaviour varies with varying Ag sputtering time. Single transient response also shows that dark current increases with increasing Ag sputtering time to 40 s.

Now as the sputtering time of Ag is further increased to 40 s, NPs formed closely packed agglomerated structure of Ag NPs schematically shown in Fig. 5c. Ag NPs makes interface with neighboring Ag NPs as shown in Fig. 5 (c and d). So in this case, hot electrons are not being able to transfer to semiconductor surface. Instead, this leads to electron tunneling coupling with its counter Ag NPs thus increase the role of dissipation by known ''non-local effects'. Thus, the 40s Ag sputtered sample has highest dark current because of the propelling metallic surfaces of the nanoparticles that approaches contact due to agglomeration and leads to the electron burrowing and gives a charge exchange pathway between surrounding Ag nanoparticles instead β-Ga$_2$O$_3$.

When illuminated by 254 nm DUV illumination these hot electrons then relax by internal decay inside a metallic Ag NPs via phonon emission which takes place under time scale of >100 ps, which leads to increase the heating of Ag NPs. This further include the heat exchange effect with nearby metallic Ag NPs, while increasing the temperature distribution in the vicinity of Ag nanoparticle-gallium oxide interface. This hot electron-phonon relaxation process can be modelled using two-temperature model (TTM) by which rate of energy transfer between electrons and phonons is given by coupled differential equation[26]:

$$C_e(T_e)\frac{dT_e}{dt} = -g(T_e - T_l) \tag{4}$$

$$C_l \frac{dT_l}{dt} = -g(T_e - T_l)$$

where $T_e$ and $T_l$ are electronic and lattice temperature of Ag NP, $c_l$ is lattice heat capacity, $C_e(T_e) = \gamma T_e$ is the temperature dependent electronic heat capacity and $g$ is the electron-phonon coupling constant. The lattice temperature (Ag NPs) can rise up extent so as to melt the Ag NP (~1000 K) if sufficient laser intensity is provided.

This heat generated may be transferred to β-Ga$_2$O$_3$ surface via phonons in time scale of 5-10 ps, which in turn increases the temperature of the β-Ga$_2$O$_3$ film. This heat transfer takes place almost instantly (few ps), which increases the resistance of β-Ga$_2$O$_3$ film considerably. Normally resistivity decrease with increasing temperature, as temperature allows more electron conduction by providing energy to them to jump over the conduction band. However, as β-Ga$_2$O$_3$ is a wide-bandgap semiconductor, so the effect of temperature to increase conduction is negligible relative to electron-electron scattering effect and hence decrement in photocurrent is observed. And as soon as the illumination is turned off, the heat gets dissipated to environment.

Although much effort has been dedicated to relieve the local heating of plasmonic non-radiative heat generation decay[12,27], this anomalous behavior can uncovered various opportunities for harnessing the heat in various potential applications which haven't been studied till now for photodetector application. For instance, welding which is used worldwide, releases radiation in DUV region (>200 nm), which is extremely dangerous for human eyes and in most cases can even bring about permanent blindness[28]. Thus the safety for them becomes crucial aspect where they constitute major population (approx. 400,000 full-time workers in the U.S in 2014). In fact in 2014, U.S. A. this matter became a profoundly serious issue when in a day statistics out of 5,702 workers 2,000 workers suffered from eye injuries [U.S. Bureau of Labor Statistics, 2014]. If we put our system in the current supply of the arena, whenever UV levels exceeds safety limits, these photocurrent drops and that will automatically turn off the power supply. Thus compared to traditional photodetectors which involves first detecting the UV radiation, which then gives feedback to shut down power supply[28], our system detect in a very early stage before any such catastrophic event takes place. In turn, this will protect more skillful labor engaged in welding where chances of occurring such event is exceptionally high.

The plasmonic induced heating was not observed for β-$Ga_2O_3$/Ag@20s because almost all the nanoparticle is in contact with gallium oxide and hot electron are generated but get transferred to gallium oxide instantaneously. The heat generated for β-$Ga_2O_3$/Ag@40s have many application to develop tunable photodetector than can work as safe-guard and act as solar-blind switch.

## CONCLUSION

Utilizing plasmonic science, we exhibited completely reversible dynamic control of Ag plasmon resonances in both segregated and associating plasmonic nanoparticles. Our outcomes demonstrate that alteration of Ag plasmonic nanoparticles permits surprisingly, reversible, and completely anomalous behaviour of the charge carriers on the photodetector. These outcomes constitute the main perception of Ag NPs arrangement in β-$Ga_2O_3$ photodetector surface. Notwithstanding giving an important model framework to exploring the changes amongst the photodetector upon Ag sputtering. Besides forcing major points of confinement on plasmon damping, the electron– phonon coupling examined here may have a part in other applications as discussed earlier. This work shows a vital utilization of plasmonics to the field of photodetector. By tuning the plasmon resonances of Ag nanoparticles suitably, it might be conceivable to populate particular electronic conditions of NP on the β-$Ga_2O_3$ surface. Therefore, our study opens up a whole new class of analysis on Ag NPs/β-$Ga_2O_3$ nanocomposites based optoelectronic devices. By comparing the conventional pristine β-$Ga_2O_3$ photodetector with plasmonic β-$Ga_2O_3$@Ag NP photodetector we have shown that the enhanced photoresponse with polarity reversion effects and anomalous behaviour of plasmonic photodetector can be well-explained and systematized by our proposed physical model to support its mechanism.

## MATERIALS AND METHODS

**Experimental Design**

The details for the growth of β-$Ga_2O_3$ thin film can be found elsewhere [16]. In brief, gallium oxide thin film was deposited using high temperature seed layer on cost effective p-Si (100) substrate. Prior to deposition, Si substrate was cleaned with soap, DI water, acetone, propanol, respectively for 30 mins in the ultra-sonification bath to remove any impurities present on the substrate. It followed the removal of $SiO_2$ present on Si with HF (5% sol). A seed layer of β-$Ga_2O_3$ (~15 nm) at $700^0C$ was deposited followed by a thin layer of β-$Ga_2O_3$ (300 nm) at $650^0C$ was deposited. The detail growth profile is discussed elsewhere [16]. During the deposition, 1 sccm $O_2$ was supplied to fill up oxygen vacancies and to minimize the defects for better optoelectronic properties of gallium oxide photodetector.

**Deposition of Ag NPs over β-Ga$_2$O$_3$ film and Characterization**

After the deposition of the β-Ga$_2$O$_3$ layer, Ag electrodes were deposited using thermal evaporator via physical mask. Thereafter a thin Ag was sputtered at low power (24 W) and high pressure (22 mtorr) to get Ag NPs decorated β-Ga$_2$O$_3$ thin film. The sputtering time was optimized to achieve the distribution of Ag nanoparticle. The β-Ga$_2$O$_3$ thin film deposited with 20 s Ag and 40 s was nomenclature as β-Ga$_2$O$_3$/Ag@20s and β-Ga$_2$O$_3$/Ag@40s, respectively. To check the morphology of the as-deposited gallium oxide and Ag decorated gallium oxide layer, Atomic Force Microscopy (AFM) was performed using Bruker AXS Multimode 8 in tapping mode. The absorption spectra were taken using Perkin Ellmer LAMBDA-950 UV-Vis-NIR spectrophotometer in the range of 200 nm – 600 nm. For the UV-Vis-NIR measurement, gallium oxide thin film was deposited on quartz substrate. The performance of the fabricated photodetector was measured using Bentham system having monochromator source. All the measurements are done in an optically tight box to avoid any optical interferences from the background.

# REFERENCES

1   Chen, H., Liu, K., Hu, L., Al-Ghamdi, A. A. & Fang, X. New concept ultraviolet photodetectors. *Materials Today* **18**, 493-502, doi:https://doi.org/10.1016/j.mattod.2015.06.001 (2015).
2   Chen, X. *et al.* Self-Powered Solar-Blind Photodetector with Fast Response Based on Au/β-Ga2O3 Nanowires Array Film Schottky Junction. *ACS applied materials & interfaces* **8**, 4185-4191 (2016).
3   Higashiwaki, M. & Jessen, G. H.    (AIP Publishing, 2018).
4   Clavero, C. Plasmon-induced hot-electron generation at nanoparticle/metal-oxide interfaces for photovoltaic and photocatalytic devices. *Nature Photonics* **8**, 95, doi:10.1038/nphoton.2013.238 (2014).
5   Atwater, H. A. & Polman, A. Plasmonics for improved photovoltaic devices. *Nature materials* **9**, 205 (2010).
6   Bohren, C. F. & Huffman, D. R. *Absorption and scattering of light by small particles*.  (John Wiley & Sons, 2008).
7   Khurgin, J. B. How to deal with the loss in plasmonics and metamaterials. *Nature nanotechnology* **10**, 2 (2015).
8   Hogan, N. J. *et al.* Nanoparticles heat through light localization. *Nano letters* **14**, 4640-4645 (2014).
9   Jacob, Z. & Shalaev, V. M. Plasmonics goes quantum. *Science* **334**, 463-464 (2011).
10  Schuller, J. A. *et al.* Plasmonics for extreme light concentration and manipulation. *Nature materials* **9**, 193 (2010).
11  Byers, C. P. *et al.* From tunable core-shell nanoparticles to plasmonic drawbridges: Active control of nanoparticle optical properties. *Science advances* **1**, e1500988 (2015).
12  Haffner, C. *et al.* Low-loss plasmon-assisted electro-optic modulator. *Nature* **556**, 483 (2018).
13  Yu, H., Pardoll, D. & Jove, R. STATs in cancer inflammation and immunity: a leading role for STAT3. *Nature Reviews Cancer* **9**, 798, doi:10.1038/nrc2734 (2009).
14  Son, J. H. *et al.* Ultrafast photonic PCR. *Light: Science & Applications* **4**, e280 (2015).
15  Liu, X. *et al.* Tunable Dipole Surface Plasmon Resonances of Silver Nanoparticles by Cladding Dielectric Layers. *Scientific Reports* **5**, 12555, doi:10.1038/srep12555 (2015).
16  Arora, K., Goel, N., Kumar, M. & Kumar, M. Ultrahigh Performance of Self-Powered β-Ga2O3 Thin Film Solar-Blind Photodetector Grown on Cost-Effective Si Substrate Using High-Temperature Seed Layer. *ACS Photonics*, doi:10.1021/acsphotonics.8b00174 (2018).
17  Alema, F. *et al.* in *Oxide-based Materials and Devices VIII.*  101051M (International Society for Optics and Photonics).
18  Bai, Z. *et al.* Self-powered ultraviolet photodetectors based on selectively grown ZnO nanowire arrays with thermal tuning performance. *Physical Chemistry Chemical Physics* **16**, 9525-9529, doi:10.1039/C4CP00892H (2014).
19  Lu, S. *et al.* Piezotronic Interface Engineering on ZnO/Au-Based Schottky Junction for Enhanced Photoresponse of a Flexible Self-Powered UV Detector. *ACS Applied Materials & Interfaces* **6**, 14116-14122, doi:10.1021/am503442c (2014).
20  Chen, H. *et al.* Ultrasensitive Self-Powered Solar-Blind Deep-Ultraviolet Photodetector Based on All-Solid-State Polyaniline/MgZnO Bilayer. *Small* **12**, 5809-5816, doi:10.1002/smll.201601913 (2016).
21  Wu, Z. *et al.* A self-powered deep-ultraviolet photodetector based on an epitaxial Ga 2 O 3/Ga: ZnO heterojunction. *Journal of Materials Chemistry C* **5**, 8688-8693 (2017).

22   Zhao, B. *et al.* An Ultrahigh Responsivity (9.7 mA W−1) Self-Powered Solar-Blind Photodetector Based on Individual ZnO–Ga2O3 Heterostructures. *Advanced Functional Materials* **27**, 1700264, doi:doi:10.1002/adfm.201700264 (2017).
23   Pratiyush, A. S. *et al.* MBE grown Self-Powered\b {eta}-Ga2O3 MSM Deep-UV Photodetector. *arXiv preprint arXiv:1802.01574* (2018).
24   Guo, D. *et al.* Zero-Power-Consumption Solar-Blind Photodetector Based on β-Ga2O3/NSTO Heterojunction. *ACS applied materials & interfaces* **9**, 1619-1628 (2017).
25   Sze, S. M. & Ng, K. K. *Physics of semiconductor devices*. (John wiley & sons, 2006).
26   Ernstorfer, R. *et al.* The Formation of Warm Dense Matter: Experimental Evidence for Electronic Bond Hardening in Gold. *Science* **323**, 1033-1037, doi:10.1126/science.1162697 (2009).
27   Hoang, C. V. *et al.* Interplay of hot electrons from localized and propagating plasmons. *Nature communications* **8**, 771 (2017).
28   Patel, N. & Dandu, P.    (Google Patents, 2017).
29   Kong, W. Y. *et al.* Graphene-β-Ga2O3 Heterojunction for Highly Sensitive Deep UV Photodetector Application. *Advanced Materials* **28**, 10725-10731 (2016).
30   Oh, S., Kim, C.-K. & Kim, J. High responsivity β-Ga2O3 metal-semiconductor-metal solar-blind photodetectors with ultra-violet transparent graphene electrodes. *ACS Photonics* (2017).
31   Qian, L.-X., Zhang, H.-F., Lai, P. T., Wu, Z.-H. & Liu, X.-Z. High-sensitivity β-Ga2O3 solar-blind photodetector on high-temperature pretreated c-plane sapphire substrate. *Optical Materials Express* **7**, 3643-3653, doi:10.1364/OME.7.003643 (2017).
32   Guo, D. *et al.* Inhibition of unintentional extra carriers by Mn valence change for high insulating devices. *Scientific reports* **6**, 24190 (2016).
33   Li, L. *et al.* Deep-ultraviolet solar-blind photoconductivity of individual gallium oxide nanobelts. *Nanoscale* **3**, 1120-1126 (2011).
34   Hsu, C.-L. & Lu, Y.-C. Fabrication of a transparent ultraviolet detector by using n-type Ga 2 O 3 and p-type Ga-doped SnO 2 core–shell nanowires. *Nanoscale* **4**, 5710-5717 (2012).
35   Cui, S., Mei, Z., Zhang, Y., Liang, H. & Du, X. Room-Temperature Fabricated Amorphous Ga2O3 High-Response-Speed Solar-Blind Photodetector on Rigid and Flexible Substrates. *Advanced Optical Materials* (2017).
36   Qian, L.-X. *et al.* Ultrahigh-responsivity, rapid-recovery, solar-blind photodetector based on highly nonstoichiometric amorphous gallium oxide. *ACS Photonics* **4**, 2203-2211 (2017).

**ACKNOWLEDGMENTS**

**Funding:** We sincerely acknowledge funding through interdisciplinary project, IIT Ropar and DRDO project (ERIP/ER/201612009/M/01/1714).

**Author details:** K.A. was in charge of the overall design of experiments, characterization and analysis of all the results presented and preparation of the manuscript under the supervision of M.K.; V.K. assisted with the Ag depositions, analysis of the results and discussion to the manuscript. All authors were included during the discussion of results and participated during the review of the manuscript.

**Conflict of interests:** The authors declare that they have no **conflict of** interests.

## SUPPLEMENTARY INFORMATION

Fig. S1. Light absorption of $\beta$-$Ga_2O_3$/Ag@20s, $\beta$-$Ga_2O_3$/Ag@40s thin films with bare $\beta$-$Ga_2O_3$ thin film.

Table S1 Comparison of $\beta$-$Ga_2O_3$ thin film photodetector at moderate bias with our work.

Fig. S2. Comparison of dark current and photo-current of all bare $\beta$-$Ga_2O_3$, $\beta$-$Ga_2O_3$/Ag@20s and $\beta$-$Ga_2O_3$/Ag@40s photodetectors in self-biased condition